\documentclass[preprint]{aastex}
\slugcomment{V7@200806025}
\shorttitle{Cyg X-1} \shortauthors{Yan, Liu and Hadrava}

\begin{document}
\title{Optical Spectroscopic Observations of Cyg X-1=HDE 226868}
\author{Jingzhi Yan\altaffilmark{1}, Qingzhong Liu\altaffilmark{1} and Petr Hadrava\altaffilmark{2}}
\altaffiltext{1}{Purple Mountain Observatory, Chinese Academy of Sciences, Nanjing , China; {\sf
jzyan@pmo.ac.cn, qzliu@pmo.ac.cn}} \altaffiltext{2}{Astronomical~Institute, Academy~of~Sciences,
Bo\v{c}n\'{\i}~II~1401, CZ-14131~Prague, Czech~Republic; {\sf had@sunstel.asu.cas.cz}}
%\affil{Purple Mountain Observatory, Chinese Academy of Sciences, Nanjing
%210008, China\\}
%
%
%\email{jzyan@pmo.ac.cn}
%\email{qzliu@pmo.ac.cn}


\begin{abstract}
We present the results of the spectroscopic observations of
HDE\,226868, the optical counterpart to the black hole X-ray binary
Cyg X-1, from 2001 to 2006. We analyze the variabilities of the two
components in the complex H$\alpha$ line: one P-Cygni shaped
component which follows the motion of the supergiant and another
emission component moving with an antiphase orbital motion relative
to the supergiant, which is attributed to a focused-stellar wind.
The results of KOREL disentangling of our spectra indicate that the
focused stellar wind is responsible for the major part of the
variability of the H$\alpha$ emission line. The emission of the
supergiant component had a small difference between the low/hard and
high/soft states, while the focused wind component became strong in
the low/hard state and weak in the high/soft state. The wind is
nearly undisturbed by the X-ray photoionization during the low/hard
state. However, during the high/soft state, the X-rays from the
compact object could decelerate the line-driven wind and result in a
high mass accretion rate, due to the effect of the X-ray
photoionization. The X-ray illuminating could also change the
temperature profile of the stellar wind and increase its
temperature, and thus decrease the H$\alpha$ emissivity of the wind,
which could explain the H$\alpha$ variabilities of Cyg X-1 during
different X-ray states.

\end{abstract}

\keywords{Stars: Binaries: Spectroscopic, Stars: Early-Type, stars: individual (HDE 226868, Cygnus
X-1), Stars: Winds, Outflows, X-Rays: Binaries}



\section{INTRODUCTION}
Cygnus X-1 was first discovered during a rocket flight observation
in 1964 \citep{bowyer65} and its optical counterpart was identified
with the supergiant star HDE 226868 \citep{bolton72, webster72}. A
5.6d orbital period was found in the optical spectroscopic
observations \citep{gies82} and the dynamical determination of the
binary components provided evidence for the existence of a black
hole in Cyg X-1. As the black hole X-ray binary, which was first
found in our Galaxy, Cyg X-1 has been extensively studied from radio
to $\gamma$-rays during recent decades. Even though the mass
function is precisely known (see \citet{gies2003}), the masses of
the donor star and the black hole are not so well constrained due to
the poorly known inclination and the evolutionary status of the
supergiant. Using the inclination of $i=35^\circ$, \citet{herrero95}
estimated the masses of the supergiant and the black hole at 17.8
$M_\odot$ and 10.1 $M_\odot$, respectively.

HDE 226868 was classified as an O9.7 Iab supergiant star
\citep{walborn73}, which shows H$\alpha$ and He {\small II}
$\lambda$4686 in emission. Although the supergiant nearly fills the
Roche lobe, the accretion is mainly via the strong stellar wind from
the donor star \citep{gies86a, gies2003}. The variabilities of the
optical lines \citep{gies86b,ninkov87} on the spectrum of HDE 226868
indicate that the distribution of the stellar wind deviates from a
spherical geometry and that an enhanced wind flow exists (``focused
stellar wind," suggested by \citet{friend82}) in the direction of
the compact object. The focused stellar wind is also revealed by the
High-Energy Transmission Grating Spectrometer aboard the Chandra
X-Ray Observatory \citep{miller2005}.

The most recent ephemeris of the 5.6d orbital period has been given
by \citet{lasala98}, \citet{brocksopp99b}, and \citet{gies2003}
according to their optical spectroscopic observations. In addition,
the 5.6d orbital period is also found in the $UBVJHK$ photometry,
X-ray and radio data of Cyg X-1 \citep{brocksopp99a}. Cyg X-1 also
shows the superorbital modulation, on a time-scale much longer than
the orbital period. A 294d period was first reported by
\citet{priedhorsky83} in X-rays and then by \citet{kemp83} in the
optical. Another $\sim$ 150d period was found by different authors
\citep{brocksopp99a,lachowicz2006} in X-ray and radio data. This
$\sim$ 150 d period may be caused by the precession of the accretion
disk around the compact object \citep{wijers99}.

A relativistic radio jet with a velocity larger than $\sim$ 0.6$c$ was detected by
\citet{stirling2001} in their radio observations of Cyg X-1, and therefore Cyg X-1 joins the group
of the Galactic microquasar. Microquasar is an X-ray binary with a pair of relativistic radio jets,
which is similar to the radio jets found in the active galactic nuclei (see \citet{mirabel99} for a
review). The jet remnant of Cyg X-1 is resolved in radio \citep{gallo2005} and optical
\citep{russell2007} observations.

Many observational characteristics of the canonical stellar black
hole candidate Cyg X-1 are considered evidence that a black hole
exisits, similar to other X-ray binary systems. These observational
features include the ultra-soft spectra, the high-energy power-law
tail above 20 keV, the spectral/temporal transition from the
high/soft state to the low/hard state and the X-ray  millisecond
variability (see \citet{tanaka95}; \citet{mcclintock2006};
\citet{remillard2006}). Most of the time Cyg X-1 is in a low/hard
state and in some years it can transit from the low/hard state to
the high/soft state, which can continue from several weeks to
several months \citep{zhang97,brocksopp99a}. Occasionally, Cyg X-1
enters an intermediate state when it fails to make a transition from
the low/hard state to the high/soft state
\citep{belloni96,malzac2006}. The radio and X-ray emission has an
anticorrelation and when the binary enters the high/soft state, the
radio emission will be decreased \citep{pooley2001}. Most
researchers believe the transition state is caused by the physical
changes of the gas around the black hole (see \citet{remillard2006}
and references therein). The physical changes in the accretion disk
are initially related to the mass loss from the supergiant
companion. However, the X-ray radiation of the disk influences the
ionization and temperature of the wind and thus also its radiation
and dynamics.

Using the method of tomographic separation\footnote{ This method of
decomposition of observed spectra into unknown spectra of two
components should be distinguished from the methods of Doppler
tomography in which the phase-locked line-profile variations are
fitted by projections of smooth distribution of delta-function
profiles in velocity space corotating with the orbital motion.}
\citep{bagnuolo91}, \citet{sowers98} showed that the H$\alpha$
profiles of Cyg X-1 observed in seasons 1985 and 1986 can be well
fitted as a superposition of a P-Cyg profile corresponding to the
(approximately spherical) stellar wind of the supergiant and a wide
emission peak radiated by the focused stellar wind, which moves in a
slightly shifted anti-phase with respect to the orbital Doppler
shift of the former component. Using the method of Fourier
disentangling \citep{hadrava95,hadrava97,hadrava04} we have proved
the same behavior in the spectra obtained at Ond\v{r}ejov
Observatory in summer 2003 shortly before, during and shortly after
one high/soft-state episode of Cyg X-1 \citep{hadrava07}.

In this article, we present our optical spectroscopic observations
of Cyg X-1 from 2001 to 2006. The preliminary results have been
introduced in a previous article \citep{yan2005}. Here we will make
a further analysis of the H$\alpha$ line profiles. Our present data
are confined to about one week in each season, meaning they are
concentrated into only slightly more than one orbital period. This
enables us to compare the solutions of disentangling in different
X-ray states, minimizing the possible influence of long-term
variations in the structure of the focused stellar wind.

In the following section, the properties of our data are described.
The method and results of the disentangling are given in Section 3.
Next, the correspondence of our observational results with the
standard model of this classical X-ray binary is studied in Section
4. Finally, the conclusions are summarized in Section 5.

\section{OBSERVATIONS}
We obtained the spectra of HDE 226868 with the 2.16m telescope at
Xinglong Station of National Astronomical Observatories, China
(NAOC), from 2001 to 2006. The optical spectroscopy with an
intermediate resolution of 1.22\,{\AA} pixel$^{-1}$ was made with a
CCD grating spectrograph at the Cassegrain focus of the telescope.
We took the red spectra covering from 5500 to 6700\,{\AA} and blue
spectra covering from 4300 to 5500\,{\AA} at different times.
Sometimes low-resolution spectra (covering from 4300 to 6700\,{\AA})
were also obtained. The journal of our observations is summarized in
Table~\ref{table}, including observational date, UT Middle, exposure
time, Julian date, wavelength range, and spectral resolution.
Orbital phase ($\phi$) is also given in Table~\ref{table} and the
ephemeris of the inferior conjunction of the companion star is
adopted from Gies et al. (2003),
  $$2,451,730.449+5.599829E\; .$$
All spectroscopic data were reduced with the IRAF\footnote{IRAF is distributed by NOAO, which is
operated by the Association of Universities for Research in Astronomy, Inc., under cooperation with
the National Science Foundation.} package. They were bias-subtracted and flat-field corrected, and
had cosmic rays removed. Helium-argon spectra were taken in order to obtain the pixel-wavelength
relations. To improve this relation, we also used the diffuse interstellar bands (DIBs) 6614 and
6379 {\AA} observed in the spectra.

\begin{center}
\begin{figure}[ht]
\centering
\includegraphics[bb=17 17 282 217, width=8cm]{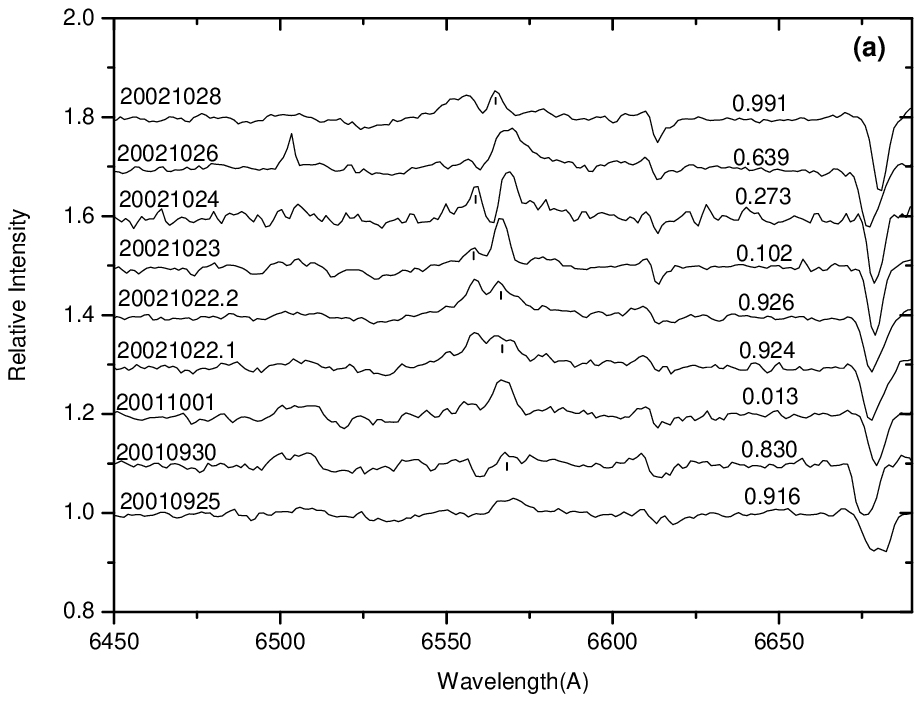}%
\includegraphics[bb=17 17 274 207, width=8cm]{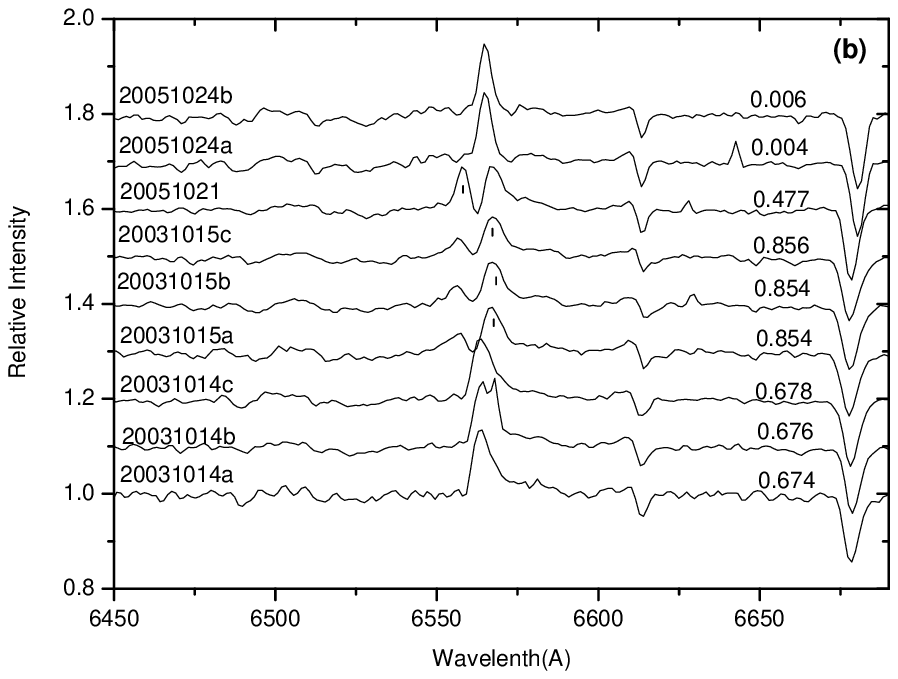}
\includegraphics[bb=17 17 274 207, width=8cm]{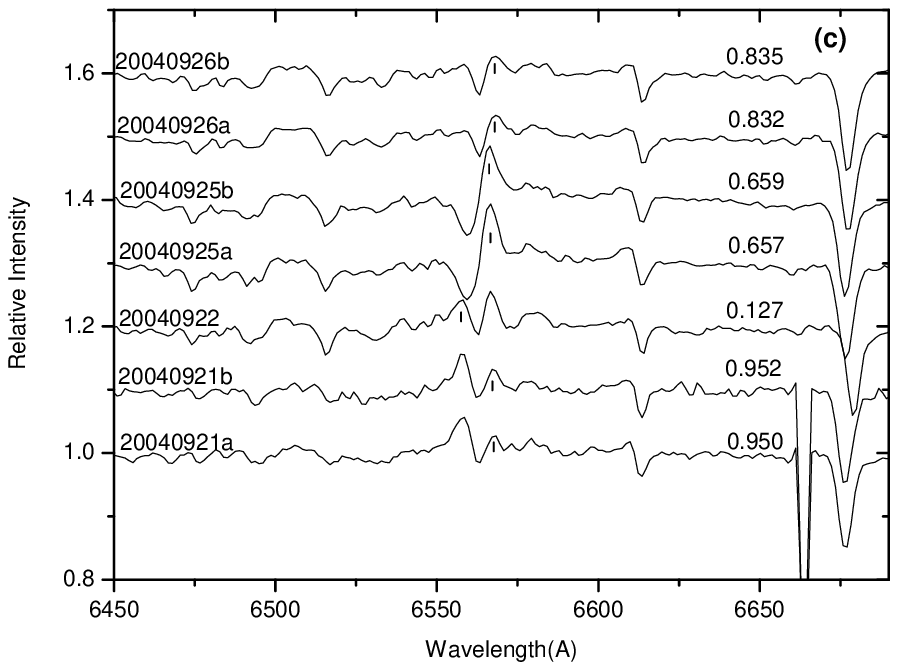}%
\includegraphics[bb=17 17 274 207, width=8cm]{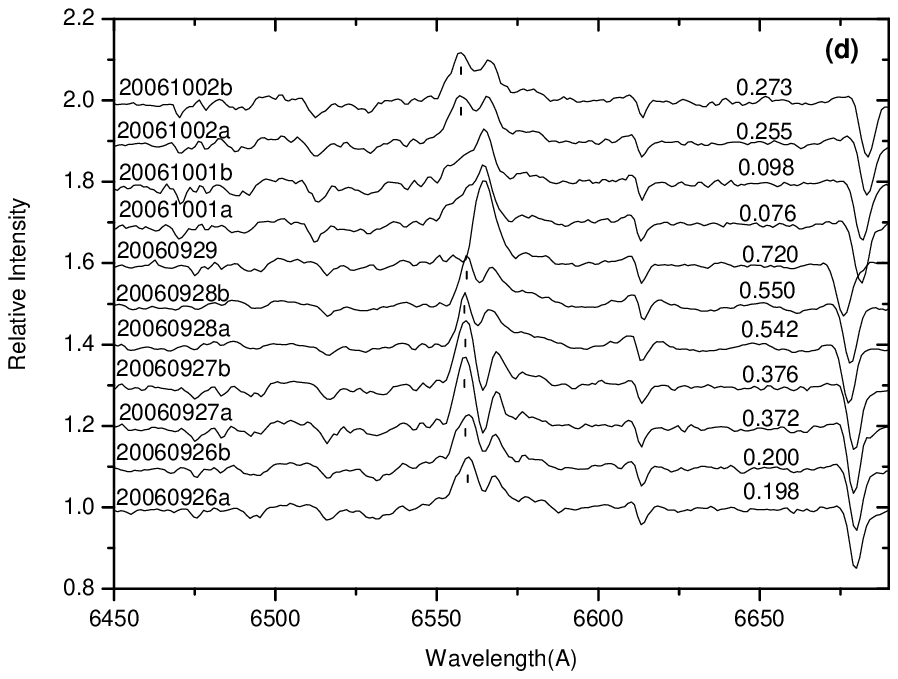}
\caption{Spectra including H$\alpha$ and He {\small I} $\lambda$6678
lines during six years' observations. The observational dates and
orbital phases are given in the left and right sides of each
spectrum, respectively: (a) H$\alpha$ spectra in 2001 and 2002, (b)
H$\alpha$ spectra in 2003 and 2005, (c) H$\alpha$ spectra in 2004,
and (d) H$\alpha$ spectra in 2006. When the H$\alpha$ shows a
double-peaked profile, the focused-wind component is marked by a
tick in the figure.} \label{figure:group}
\end{figure}
\end{center}

The higher resolution red spectra obtained from 2001 to 2006 are
shown in Figure~\ref{figure:group}. The corresponding observational
dates and orbital phases calculated according to the above given
ephemeris are written on the left and right sides of each spectrum,
respectively. The spike on the left part of the spectrum on 2002
October 26 may be caused by a cosmic hot point and the dips on the
spectra of 2004 September 21 are caused by bad pixels on the CCD.
Most of the H$\alpha$ lines show a double-peaked profile with a
central absorption. Single-peaked H$\alpha$ lines are observed in
our spectra of 2003 October 14 ($\phi$=0.67), 2005 October 24
($\phi$=0.004) and 2006 September 29 ($\phi$=0.72). Obvious P-Cygni
H$\alpha$ lines are observed in our spectra in some phases only in
2004 (Figure~\ref{figure:group}(c)).

For a comparison, we have also used the spectra obtained with the
700mm camera of the Coud\'{e} spectrograph at the 2.05m telescope of
the Ond\v{r}ejov observatory (the Astronomical Institute of the
Czech Academy of Sciences). These spectra covering the region
6260--6760 {\AA} with a resolution of approximately
0.25\,{\AA}\,pixel$^{-1}$ are included in a study by
\citet{gies2008}, to which we refer for details.

\section{ANALYSIS AND RESULTS}
\subsection{The Equivalent Width Evolution of H$\alpha$ and the X-ray Activity}
The equivalent width (EW) of the complex H$\alpha$ line (emission
and absorption) has been measured selecting a continuum point on
each side of the line and integrating the flux relative to the
straight line between the two points using the procedures available
in IRAF. The measurements were repeated five times for each spectrum
and the error estimated from the distribution of values obtained.
The typical error for H$\alpha$ measurements is within 10\%. This
error arises due to the subjective selection of the continuum. The
results of H$\alpha$ EWs are listed in Table~\ref{table}.

\begin{center}
\begin{figure}[ht]
\centering
\includegraphics[bb=11 11 213 283, width=10cm]{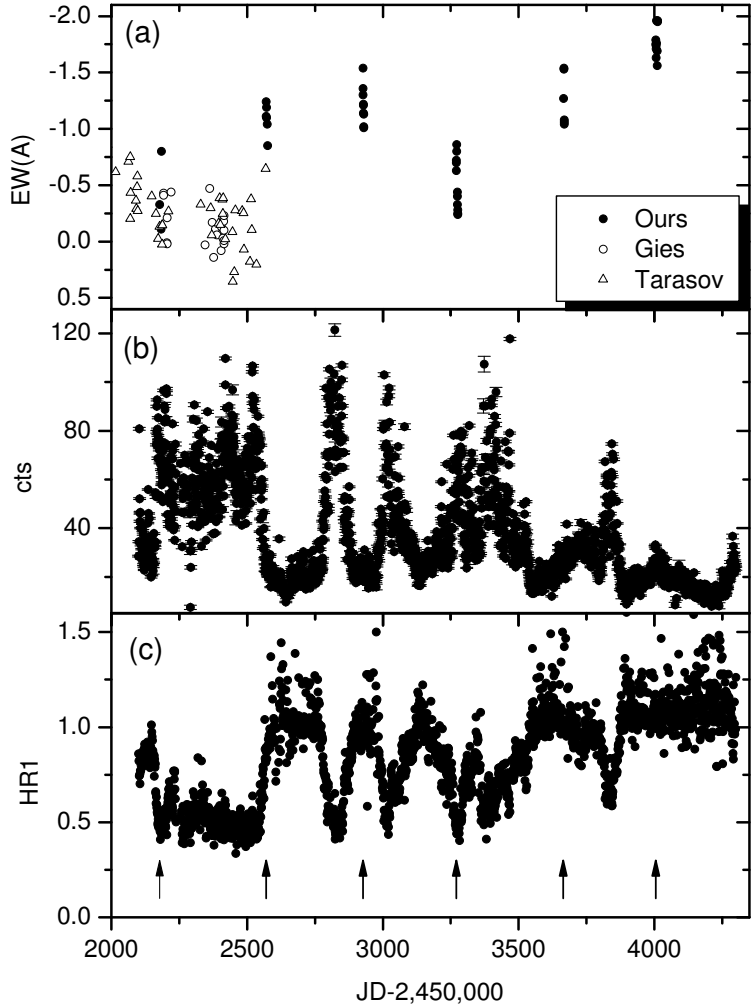}
\caption{(a) H$\alpha$ EWs during six-year observations (filled
circles). Data from \citet{gies2003} (open circles) and
\citet{tarasov2003} (open triangles) are shown. (b) The
one-day-averaged RXTE/ASM count rates of Cyg X-1 in 1.5-12 keV. (c)
The hardness ratio in the soft X-ray band of RXTE/ASM, (3-5
keV)/(1.5-3 keV). The arrows on the bottom of the panel correspond
to the starting time of each observational run. } \label{figure:ew}
\end{figure}
\end{center}

The top panel of Figure~\ref{figure:ew} shows the variability of the H$\alpha$ EW from 2001 to 2006
as a function of time. In addition to the data obtained in our observational program, the combined
data sets of \citet{gies2003} and \citet{tarasov2003} are also included in the figure. For a
comparison, the middle panel of Figure~\ref{figure:ew} gives the RXTE/ASM one-day averaged counter
rate in the 1.5-12 keV band and the hardness ratio (HR1) of the soft X-ray radiation, (3-5
keV)/(1.5-3 keV) is plotted in the bottom panel. The arrows in this panel correspond to the
starting date of each observational run.

Figure~\ref{figure:ew} shows that our 2001 and 2004 observational
runs were done in a high/soft state while 2003, 2005, and 2006
observational runs in a low/hard state. The observations in 2002
were during a transitional state from high/soft to low/hard. The EW
of H$\alpha$ is relatively low in the high/soft state and strong in
the low/hard state. This phenomenon has been discussed in detail by
\citet{gies2003} and \citet{tarasov2003}. While the H$\alpha$ in our
spectra of 2001 nearly lost its emission signature, the emission
level of H$\alpha$ in the 2006 observations is the strongest among
our six observational runs. Because the season 2001 is poorly
covered by observations, we chose the observational runs 2004 and
2006 to represent the high/soft and low/hard states, respectively,
in our study of line-profile variability.

\subsection{The Profile Variability of H$\alpha$ Emission Line}
The top panel of Figure~\ref{figure:2006} shows the H$\alpha$
profiles during our 2006 observational run. The exposures are
depicted in ascending order according to the orbital phase. The
observational date and orbital phase are marked on the left and
right sides of each spectrum, respectively. We have recalibrated the
wavelength scale of each spectrum according to the position of the
DIB 6614\,{\AA}. It can be seen from this figure that, for most of
the time, the H$\alpha$ of Cyg X-1 has a double-peaked profile, one
peak formed in the supergiant and the other in the focused stellar
wind between the system components \citep{sowers98,gies2003}. As in
many other supergiants, its intrinsic H$\alpha$ emission is due to
the powerful stellar wind \citep{puls96} and it forms the
red-shifted emission wing of the P-Cygni line profile that follows
the orbital motion of the star. In some phases (e.g., 0.720, which
is close to one extreme of the radial velocities), this emission
merges with the emission of the focused wind moving almost in an
anti-phase and they form a single bright peak. In some other phases,
the emission of the focused wind can fill the absorption part of the
P-Cygni component formed in the supergiant. Thus we can detect a
characteristic P-Cygni H$\alpha$ line in the spectrum of Cyg X-1
only in some phases of the high/soft state, when the focused-wind
emission is weak. The bottom panel of Figure~\ref{figure:2006} shows
a gray-scale map of the H$\alpha$ profiles in 2006.

Following \citet{sowers98}, \citet{gies2003} used the tomographic
separation algorithm to decompose the H$\alpha$ line profile into
the two components for chosen combinations of the radial velocity
semiamplitude $K_{em}$ and phase shift $\phi$ of the focused wind
(orbital parameters of the supergiant being fixed from nonemission
lines). Minimizing the residuals in the two-parameter space, they
found that the focused wind component has a maximum radial velocity
$K_{em}=218\pm$30\,km\,s$^{-1}$ near the orbital phase
$\phi_0=0.79\pm$0.04. The radial velocity curves of the two
components are plotted in Figure~\ref{figure:radial}. The positions
of the focused-wind component of the H$\alpha$ line based on this
solution are also marked by ticks in Figure~\ref{figure:group}.

\subsection{Disentangling the H$\alpha$ Line Profile}

The methods of disentangling fit the observed spectra as a
superposition of several components with simultaneously optimized
orbital parameters (and/or some other free parameters). Usually, the
emission-line objects are variable on a time scale shorter than the
orbital period. This creates problems for disentangling, tomographic
separation, Doppler imaging or any other similar method that
requires observations spanning a time interval at least of this
order (unless the variability is properly involved in the model of
component spectra). However, there is always a chance that such
methods can reveal a mean behavior of the object treating the rapid
variability as a noise.

The results by \citet{sowers98} and \citet{gies2003} indicate that
this assumption can work well in the case of Cyg X-1. To decompose
the spectra of the supergiant and the focused stellar wind, we thus
use the KOREL code \citep{hadrava04} for Fourier disentangling
\citep{hadrava95}, which enables us to take into account and to
resolve instantaneous changes in strength of lines of each component
\citep{hadrava97}. In the preliminary results for the Ond\v{r}ejov
data \citep{hadrava07} the He\,I line at 6678\,{\AA} has been
disentangled into a weak telluric contribution and a pure absorption
line of the supergiant, which confirms within the observational
errors the orbital parameters, found by classical measurements
(e.g., $K_{1}=71.9$\,km\,s$^{-1}$ compared to $75.6\pm.7$ by
\citet{gies2003}, or $73.0\pm.7$ by \citet{gies2008}). The H$\alpha$
line has been disentangled into three components corresponding to
the supergiant, the focused wind, and the telluric water vapor
lines, which are relatively strong in some Ond\v{r}ejov spectra.
Because, unlike the orbital velocities of the supergiant and the
black hole, the mean velocity of the focused stellar wind is not
perpendicular to the line joining the components of the binary, the
option of KOREL to disentangle up to five components in a
hierarchical structure has been used to identify the focused wind
with a component of a second close pair corotating with the
supergiant-black hole pair with identical period but with a shifted
phase. The results showed the P-Cyg profile moving with the
supergiant and a broad emission peak of the focused wind. The
orbital parameters of the P-Cyg profile converged in the
disentangling to values consistent with the He-line solution (e.g.,
$K_{1}=71.3$\,km\,s$^{-1}$). The variations of its strength were
relatively small (of the order of 0.1); moreover the EW of this
component is also small, because the red emission wing nearly
compensates the blue absorption wing. For the broad emission
component we found a semiamplitude of $K_{em}=60.8$\,km\,s$^{-1}$,
which agrees better with the value 68\,km\,s$^{-1}$ by
\citet{sowers98} rather than the value obtained by \citet{gies2003};
compare their figures 5 and 6, respctively. The broad minima of the
spectra residuals in the parameter space are due to the large width
and the variability of the focused-wind emission component, and the
position of the deepest point may be influenced by long-term changes
in the circumstellar matter as well as by the random sampling by the
observations. The line-strength factor derived by disentangling of
the Ond\v{r}ejov spectra for the focused wind is significantly
higher for this component (reaching a value around +1 at the initial
and final low/hard H$\alpha$ emitting states). Because the absolute
value of EW of the focused wind (which is negative) is higher, the
variability of this component is responsible for the major part of
the enhancement of the H$\alpha$ emission in the low/hard state. In
the disentangling of the Ond\v{r}ejov data, a small part of the
H$\alpha$ emission appeared in the telluric spectrum, which was left
arbitrary to test the behavior of the solution. Because the annual
changes of the heliocentric radial velocity corrections are
appreciably smaller compared with the amplitudes of both the
supergiant and the focused-wind disentangled components, this part
of the emission belongs to circumstellar matter which is in the mean
at rest with respect to the center of mass of the binary system.

In disentangling the NAOC spectra we struggled with two instrumental
obstacles. One was the smaller spectral resolution and the other was
the unreliability of the dispersion curve, probably caused by
insufficient stiffness of the Cassegrain spectrograph. Due to the
former, no telluric lines can be seen in these spectra. Fortunately,
it seems that they are not as strong as they are in some
Ond\v{r}ejov spectra, in which they can be distinguished even after
a smoothing. Consequently, they do not need to be disentangled to
clean the stellar spectra, but at the same time they cannot be used
to check or improve the wavelength scale, as described by
\citet{hadrava06}. We thus tried to use the above-mentioned DIBs
6614 and 6379 {\AA} for this purpose. We measured their positions in
the Ond\v{r}ejov spectra first to find their mean central
wavelengths in Cyg X-1 and then in the NAOC spectra to get the same
values by a linear transformation of the wavelength scale in each of
these spectra. For the measurement we used a single-component
disentangling by KOREL in the option of free radial velocities in
each exposure (e.g., \citet{hadrava04}). This procedure improved the
wavelength scale to some extent, yet a solution of the He\,I line
6678\,{\AA} made in order to check the reliability of the improved
wavelength scale revealed errors in some exposures up to about
60\,km\,s$^{-1}$ and similarly the disentangling of H$\alpha$ did
not provide satisfactory results. The insufficient precision of the
results may be caused by the weakness of the DIB 6379\,{\AA}, which
was still measurable in the Ond\v{r}ejov data, but too shallow and
wide in the lower-resolution data from the NAOC.

We thus chose an alternative way to disentangle the H$\alpha$
profiles in spite of unreliable wavelength scale. First we
disentangled both seasons 2004 and 2006 together with the
Ond\v{r}ejov data for the three components (the supergiant, the
focused wind and the telluric lines) in the option of free
velocities. The velocities and line strengths were kept fixed for
the Ond\v{r}ejov exposures from their previous standard solution.
This solution was thus used as a template to which radial velocities
and line strengths of each NAOC exposure were adjusted. The line
strengths were also converged by the simplex method (instead of
direct least squares which would also change the Ond\v{r}ejov
values) for the first two components in these exposures and
prescribed to a large negative value for the telluric component to
diminish it in the NAOC exposures. (The difference in spectral
resolution of these two data sets proved to be unessential in these
wide profiles.) The results of this solution were used as the
initial approximation for two component solutions (without the
telluric lines) of free radial velocities and line strengths (now by
the direct calculation) independently for both seasons 2004 and 2006
without the other data. These solutions thus cannot yield correct
radial velocity curves for each component and proper wavelengths of
the disentangled spectra, but they give correct differences between
the radial velocities of the two components, their line profiles,
and the line strengths.

The disentangled H$\alpha$ profiles and the fit of the input spectra
for our 2004 and 2006 observations are plotted by the standard
graphical KOREL output in Figure~\ref{figure:korelnew} at the left
and right panels, respectively. These spectra span from 6540\,{\AA}
to about 6596\,{\AA}; however, the shift of each profile up to
several {\AA} is uncertain due to the above-explained problems. It
can be seen (at the upper 7 and 11 curves, respectively.) that the
agreement of the fit with the observation is quite good (some
differences appear for the fifth and sixth exposures from the top in
the right panel only, i.e. for 2006 September 28).

The bottom two curves show the disentangled P-Cyg profile of the
supergiant (the higher of these two curves) and the mean
disentangled emission profile of the focused stellar wind (the very
bottom curve) in each season. The tops of the emission wings of the
P-Cyg profiles are 1.06 and 1.09 of the level of continuum in 2004
and 2006, resp. This indicates a 50\% increase above the continuum
in the low/hard state in 2006; however, this result deserves a
confirmation by more extensive observations, because the small
change representing 3\% of the continuum may be influenced only by
uncertainties of the continuum intensity (e.g., due to differences
in phase sampling). In any case, this change in the P-Cyg component
is negligible in comparison with the overall enhancement of
H$\alpha$ emission in the low/hard state. The absorption wing is
shallower in 2004 than in 2006 with the deepest point at levels 0.93
in 2004 and 0.83 in 2006. It also can be seen that the decrease of
the emission intensity toward the higher velocities is somewhat
slower in 2006 and similarly the depth of absorption is more
pronounced for higher velocities in the 2006 low/hard state. The EW
0.54\,{\AA} of the absorption part of the P-Cyg profile in 2004 is
canceled (with precision almost 10$^{-3}$\,{\AA}) by EW
--0.54\,{\AA} of the emission part of the profile. In 2006, the EW
of P-Cyg absorption is +1.910\,{\AA} and emission --1.916\,{\AA}.
Note that these EWs refer to the disentangled mean seasonal profile,
for which the noise is decreased and continuum is fitted to an
extended spectral region. Consequently, their errors are smaller
than those given in Table~\ref{table} for individual exposures
(which are in the mean $\pm 0.03$\,{\AA} and $\pm 0.05$\,{\AA} in
2004 and 2006, respectively.). The continua of disentangled
components may suffer from some complementary distortions induced by
errors in rectification of the input spectra. The number of
significant digits given here is to show that emission and
absorption EWs of the P-Cyg component cancel each other out in both
states within the precision of our data.

More remarkable difference is found between the disentangled
profiles of the emission of the focused wind: its maximum is 0.056
in 2004, and 0.265 in 2006 in the units of the continuum level of
the supergiant (a possible continuum of the focused wind cannot be
disentangled, but it must be negligible compared with the continuum
of the supergiant). Similarly, the mean EWs of the focused-wind
emission are --0.33\,{\AA} in 2004 and --1.78\,{\AA} in 2006.

The line-strength factors, like the EWs, seem to show some
phase-locked variations, which could reveal the geometry of the
system (i.e., the distribution of the focused stellar wind in the
space between the components) and eventually also the anisotropy of
the deeper layers of the stellar wind in the upper atmosphere of the
supergiant. However, the amplitude of these changes is comparable to
the observational scatter and the phase coverage of our data is not
yet sufficient, so we postpone this problem to our next study.

Our results generally confirm that it is the contribution of the
focused wind, and not the P-Cyg profile formed in the root of the
supergiant's wind, that is responsible for the major part of the
H$\alpha$ emission enhancement in the low/hard state compared with
the high/soft state. The physics of this process will be discussed
quantitatively in the following section.

\section{DISCUSSIONS}
\subsection{The X-ray Excited Wind Model}

The wind from isolated O and B stars is accelerated to a speed of
approximately three times the escape velocity from the surface of
the star by the force arising from the absorption and scattering of
the photospheric continuum radiation in the ultraviolet resonance
lines of abundant ions in the wind \citep{castor75}. The radiation
driving force produces a velocity profile of
\begin{equation}
v(r)=v_{\infty}(1.0-R_*/r)^{\beta}
\end{equation}
in a steady state, where $v(r)$ is the wind velocity at a distance
of $r$ from the center of the star, $v_{\infty}$ is the terminal
velocity of the stellar wind, $R_*$ is the radius of the supergiant,
and $\beta$$\approx$0.8 \citep{friend86,pauldrach86}. The
gravitational effect of the compact companion in a massive X-ray
binary system induces a stream of enhanced wind (focused stellar
wind) in the line from the supergiant to the compact object. The
level of the density enhancement in the focused wind is less than a
factor of 2 as derived by \citet{haberl89}, while \citet{blondin91}
thought that the density in the focused wind was 20-30 times the
ambient wind density. The presence of the focused wind in the
massive X-ray binary could greatly enhance the mass accretion rate
of the compact object in the system. Meanwhile, the X-ray radiation
from the compact object can strongly influence the dynamics of the
wind via X-ray heating and photoionization
\citep{blondin90,blondin91}.

The acceleration of the wind can be inhibited by the X-ray
photoionization which can enhance the degree of ionization in the
stellar wind. Thus, the deceleration of the wind due to the
photoionization will greatly enhance the mass accretion of the
compact object, which is a sensitive function of the wind velocity
law, $\dot{M}_{acc}$$\propto$$v^{-4}$, and lead back to a higher
X-ray luminosity. The temperature and ionization state of the
stellar wind depend only on the ``ionization parameter"
\citep{kallman82},
\begin{equation}
\xi=L_x/n_pr_x^2,
\end{equation}
where $L_x$ is the X-ray luminosity of the compact object, $n_p$ is
the proton number density of the wind, and $r_x$ is the distance to
the X-ray source. When $\xi$$\geq$10$^2$ ergs cm s$^{-1}$, the
stellar wind will be strongly affected by X-ray ionization and when
$\xi$$\geq$10$^3$ ergs cm s$^{-1}$, X-ray heating will affect the
dynamics of the wind \citep{blondin90}. To estimate the ionization
parameter, let us suppose the wind of the supergiant to be
spherically symmetric with a constant mass-loss rate,
$\dot{M}$=4$\pi$$r^2$$\mu$$m_p$$n_p(r)$$v(r)$, where $\mu$ is the
mean atomic weight and $m_p$ is the mass of the proton. Then the
ionization parameter in terms of the position and the stellar wind
parameter can be derived \citep{sako99},
\begin{equation}
\xi(r,r_x)=4.3\times10^2\frac{(L_x)_{36}(v_{\infty})_8}{\dot{M}_{-7}}(\frac{r}{r_x})^2(1-\frac{R_*}{r})^{\beta},
\label{equation:ionization}
\end{equation}
where $(L_x)_{36}$ is the X-ray luminosity in a unit of $10^{36}$ ergs s$^{-1}$, $(v_{\infty})_8$
is the terminal velocity in a unit of 10$^8$ cm s$^{-1}$, and $M_{-7}$ is the mass loss of the
supergiant in a unit of 10$^{-7}$ $M_{\odot}$ yr$^{-1}$.

Using the transformation formulae of \citet{zdziarski2002}, we can
convert the RXTE/ASM one-day averaged counts in three bands (1.5-3,
3-5, and 5-12 keV) to the energy units. Adopting a distance of 2.5
kpc \citep{ninkov87} to the source, we can get the X-ray luminosity
of Cyg X-1 in different X-ray states. The mass-loss rate of the O9.7
supergiant is calculated by the formulae given by \citet{howarth89},
$\dot{M}$=2.0$\times$10$^{-6}$M$_{\odot}$ yr$^{-1}$. All parameters
of Cyg X-1 are listed in Table~\ref{parameter}. The ionization
parameters during our 2001, 2003, 2004 and 2006 observational runs
are calculated according to Equation~(\ref{equation:ionization}) for
two distances $r_x$=21$R_{\odot}$ and 23$R_{\odot}$ from the black
hole and they are listed in Table~\ref{ionization}. We also plot the
contours of constant $\xi$ ($\xi$=10$^2$) for Cyg X-1 in different
X-ray states in Figure~\ref{figure:contour}, where the coordinate
origin is in the center of the supergiant, the thick circle is the
surface of the supergiant, and the intersection of the two black
dashed lines is the boundary of the supergiant's Roche lobe.

\subsection{The Mass Accretion Rate in Different X-Ray States}
The X-ray activity of the compact object is closely related to the
mass accretion rate onto it. Since the supergiant component of Cyg
X-1 has a negligible change in the H$\alpha$ line between the
low/hard and high/soft X-ray states, as discussed in the previous
section, the mass-loss rate from the supergiant does not have an
obvious change between these two types of X-ray states. One possible
factor causing the enhancement of the mass accretion rate during the
high/soft state may be the X-rays from the compact object.
Figure~\ref{figure:contour} indicates that the ionization parameter
$\xi$ has a value of about 100\,ergs\,cm\,s$^{-1}$ at the position
of $r_{x}$=21$R_{\odot}$ (in the line from the compact object to the
supergiant) during the high/soft state, while it only has a value of
11\,ergs\,cm\,s$^{-1}$ during the low/hard state. Consequently, the
focused wind may be photoionized by the X-ray during the high/soft
state, which leads to the decrease of the radiative pressure exerted
in lines by the supergiant's own radiation, and hence also the
velocity of the gas will be decreased. The slower wind velocity will
greatly enhance the mass accretion of the compact object and a high
X-ray luminosity will be observed during the high/soft state. During
the low/hard state, the X-ray radiation is not strong enough to
affect the focused wind, hence Cyg X-1 has a relatively low-mass
accretion rate.

A related question we should discuss is which kind of mechanism causes Cyg X-1 to transit from the
low/hard state to the high/soft state. \citet{done2002} suggested that the disk instability
mechanism (DIM; \citet{lasota2001}) triggered the X-ray outburst and then the X-ray irradiation
photoionized the hydrogen in the wind of the primary and enhanced the mass accretion onto the
compact object. The occurrence of the X-ray outburst does not have any periodicity and it may
happen at any instant in the course of time. The duration of the outburst may be connected with the
interaction between the X-rays from the compact object and the wind of the supergiant. In general,
it can keep several weeks or even several months (see Figure~\ref{figure:ew}).

\subsection{The X-Ray Irradiation Effects on the H$\alpha$ Emission}
X-ray heating could affect the dynamics of the wind from the
supergiant. According to the plane-parallel model by \citet{wu2001},
a temperature-inversion layer will be formed in the stellar wind
under the X-ray illumination. The temperature profile of the
temperature-inversion layer is determined by the soft and hard X-ray
irradiation. A strong temperature-inversion layer can be formed
provided that the incident X-rays are soft and their angle of
incidence is approaching grazing incidence. The more penetrating
hard X-rays tend to heat up the deeper layers of the stellar wind.

During the high/soft state, the increased soft X-ray heating will
increase the temperature of the ambient wind gas and then decrease
the H$\alpha$ emissivity, which is proportional to $T^{-1.2}$
\citep{richards98}. The X-ray heating is more effective in the
position near the X-ray source so the focused wind component in the
Cyg X-1 is more easily affected by the X-ray radiation from the
compact object. Thus, the focused-wind component in the H$\alpha$
line becomes weak during the X-ray high/soft state. During the
low/hard state, the temperature profile of the focused wind is
undisturbed by the X-ray irradiation and thus it has a strong
emission in the H$\alpha$ line. Compared with the focused wind, the
supergiant is farther from the X-ray source and its spherical wind
is rarely influenced by the X-ray heating. However, when the X-ray
luminosity is very high, the wind near the supergiant also can be
affected (see Figure~\ref{figure:contour}). For the high/soft state
during our 2004 observational run the outer parts of the supergiant
stellar wind were more ionized and thus unable to absorb and re-emit
in H$\alpha$. The terminal velocity of the wind also was smaller.
For the low/hard state during our 2006 observational run, the soft
X-rays from the compact object have almost no influence on the
supergiant stellar wind. This could explain the small differences
between the shapes of disentangled profiles in the two states in
Figure~\ref{figure:korelnew}. When the soft X-ray flux was very
strong, the wind near the supergiant could also be influenced by the
X-ray photoionization. Thus the system might lose its emission
feature both in the supergiant and the focused-wind components. This
scenario is consistent with the weak emission H$\alpha$ line in our
2001 spectra (on the bottom of Figure~\ref{figure:group}(a)).

During our 2004 observation, Cyg X-1 was also in a high/soft state,
but the X-ray intensity was not so strong as that in our 2001
observations. Therefore, the focused-wind component in the H$\alpha$
line was at a lower emission level, while the supergiant component
had a small change relative to the case in the low/hard state. Then
we could observe an obvious P-Cygni H$\alpha$ profile during our
2004 observations. The spectra of \citet{gies2003} and
\citet{tarasov2003} also indicate that the P-Cygni H$\alpha$ profile
often appeared in the X-ray high/soft state. During the low/hard
state, we rarely observed the P-Cygni structure in the complex
H$\alpha$ line of Cyg X-1, because the focused-wind H$\alpha$
component is strong enough to fill the absorption part of the
P-Cygni line formed in the wind of the supergiant.

\subsection{The X-ray Flare During Our 2006 Observations}

The H$\alpha$ EWs in our 2006 observational run are stronger than
those in other runs (see Figure~\ref{figure:ew}).
Figure~\ref{figure:group}(d) shows that the focused-wind component
in our 2006 H$\alpha$ spectra becomes very strong while the
supergiant component is still in its normal emission level. Since
the supergiant component did not have a significant change during
our 2006 observations, it is inappropriate to explain the enhanced
emission in the focused-wind H$\alpha$ component using the increased
mass-loss rate from the supergiant. Figure~\ref{figure:ew} indicates
that there is a small X-ray flare around the 2006 observational run.
This X-ray flare was also detected by Swift/BAT in the 15-20 keV
high-energy band. This phenomenon has been reported by
\citet{albert2007}. Figure~\ref{figure:flare} shows the H$\alpha$ EW
evolution during our 2006 observations and the light curves of the
flare detected by RXTE/ASM and Swift/BAT, respectively. The dashed
lines correspond to the beginning and the ending times of our 2006
observations.

This flare differs from the X-ray outburst during the high/soft
state. It has a relatively low X-ray flux and a hard X-ray spectrum.
The TeV emission also was detected during the flare by
\citet{albert2007} and they suggested that the flare was caused by
the interaction between the jet and the stellar wind. Our
observations were carried out on the decline phase of this flare.
The weak soft X-ray emission during the flare is insufficient to
form a prominent temperature-inversion layer in the focused-wind of
Cyg X-1 and therefore a strong focused wind H$\alpha$ component was
observed. However, the hard X-ray component during the flare could
heat up the deeper layer of the focused wind and an extra H$\alpha$
emission could be detected during our spectroscopic observations.
This can explain the strongest H$\alpha$ emission during our 2006
observations.

Another similar strong H$\alpha$ line with an EW of --2.315\,{\AA}
was also observed by \citet{tarasov2003} on MJD 5,0941.5449, when a
small X-ray flare (low/hard state) was also detected by RXTE/ASM in
the 1.5-12 keV band (see the inset in the top panel of
Figure~\ref{figure:flare}).

\section{CONCLUSIONS}

We present and analyze our optical spectroscopic observations of Cyg
X-1 from 2001 to 2006. Combined with the RXTE/ASM X-ray
observations, we make the following findings about this classical
galactic black hole X-ray binary:
\begin{enumerate}
\item We confirm that the H$\alpha$ line shows two components:
a P-Cygni profile moving with the radial velocity curve of the
supergiant and a focused stellar wind component that moves with an
approximately anti-phase orbital motion relative to the supergiant.
The superposition of the two components forms the complex H$\alpha$
profiles.

\item The results of KOREL disentangling the H$\alpha$ spectra during our 2004 and 2006
observations indicate that the focused-stellar wind is responsible
for the major part of the H$\alpha$ variability between different
X-ray states. The focused wind component becomes strong during the
X-ray low/hard state and weak during the high/soft state. The
photoionization and heating of the X-ray photons from the compact
object may affect the ionization state and dynamics of the wind from
the supergiant. During the high/soft state, the X-ray
photoionization could decelerate the gas in the focused stellar wind
and result in an increasing mass accretion rate. During the low/hard
state, the X-ray is not strong enough to influence the wind
ionization state and the compact object has a low-mass accretion
rate. The X-ray illumination can form a temperature-inversion layer
in the stellar wind. During the high/soft state, the soft X-rays
acting onto the focused stellar wind could increase its temperature
greatly and thus decrease the H$\alpha$ emissivity. This could
explain the variability of the focused-wind component in H$\alpha$
during different X-ray states. The strong soft X-ray emission during
the high/soft state could also ionize the outer parts of the
supergiant winds and render it unable to absorb and re-emit in
H$\alpha$. This scenario is consistent with the small differences
between the shapes of disentangled P-Cygni components in the 2004
and 2006 spectra. During our 2001 observations, the wind near the
supergiant was also affected by the strong X-ray emission and an
extremely weak H$\alpha$ line was observed.

\item The H$\alpha$ lines in our 2006 observations are very strong.
We interpret this as the result of the low irradiation of the focused wind by the X-ray photons.
% formed in the interaction between the jet and the ambient wind.
The weak incident soft X-rays during the flare could not disturb the temperature profile of the
focused stellar wind, while the hard X-rays could heat up the deep layer of the wind and an extra
H$\alpha$ emission could be observed.

\end{enumerate}

\begin{center}
\begin{figure}[ht]
\centering
\includegraphics[bb=58 32 387 386, width=8cm]{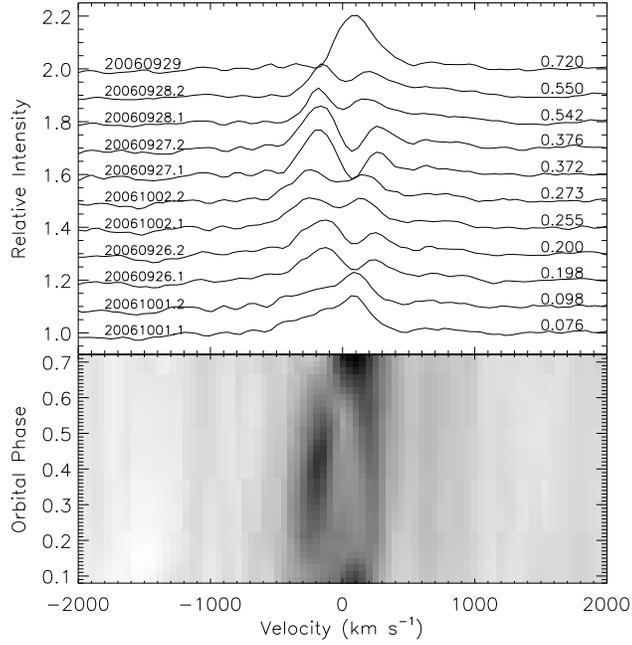}
\caption{Top: H$\alpha$ profiles during the 2006 observations,
arranged in ascending order according to the orbital phase. The
observational dates and orbital phases are given on each side of the
H$\alpha$, respectively. All spectra have had the continuum level
normalized and offset vertically to allow direct comparison. Bottom:
the gray-scale map of the H$\alpha$ spectra in 2006. The y-axis
corresponds to the orbital phase and the gray intensity is scaled
between 0.986 (white) and 1.17 (black).} \label{figure:2006}
\end{figure}
\end{center}

\begin{center}
\begin{figure}[ht]
\centering
\includegraphics[bb=25 15 288 270, width=8cm]{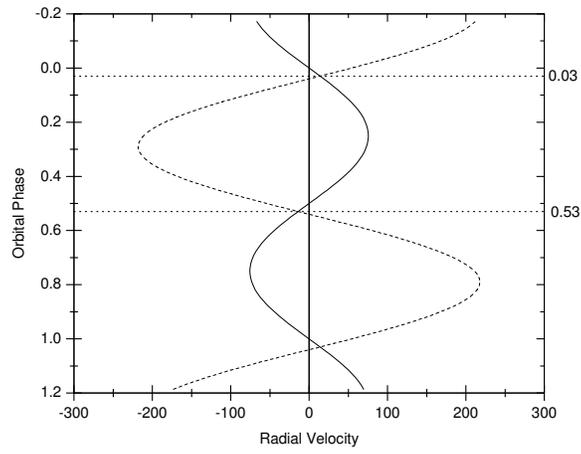}
\caption{Radial velocity curves of the supergiant (solid line) and
focused-wind (dotted line)
  components in the H$\alpha$ line adopted from \citet{gies2003}. The two components have a comparative
  velocity around orbital phases $\phi$=0.03 and 0.53. } \label{figure:radial}
\end{figure}
\end{center}

\begin{center}
\begin{figure}[ht]
\centering
\includegraphics[bb= -10 -10 510 510, width=8cm]{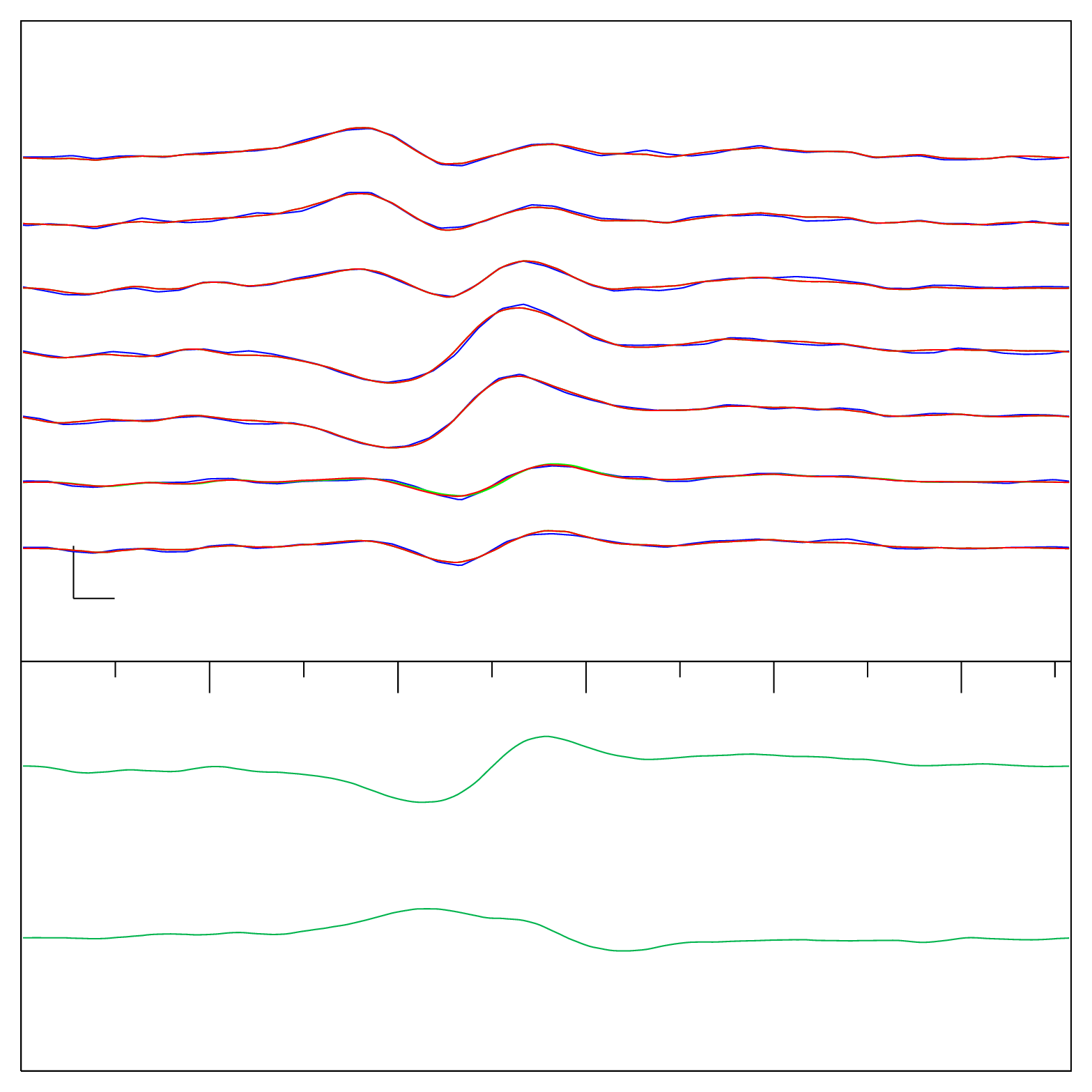}%{sc.eps}%
\includegraphics[bb= -10 -10 510 510, width=8cm]{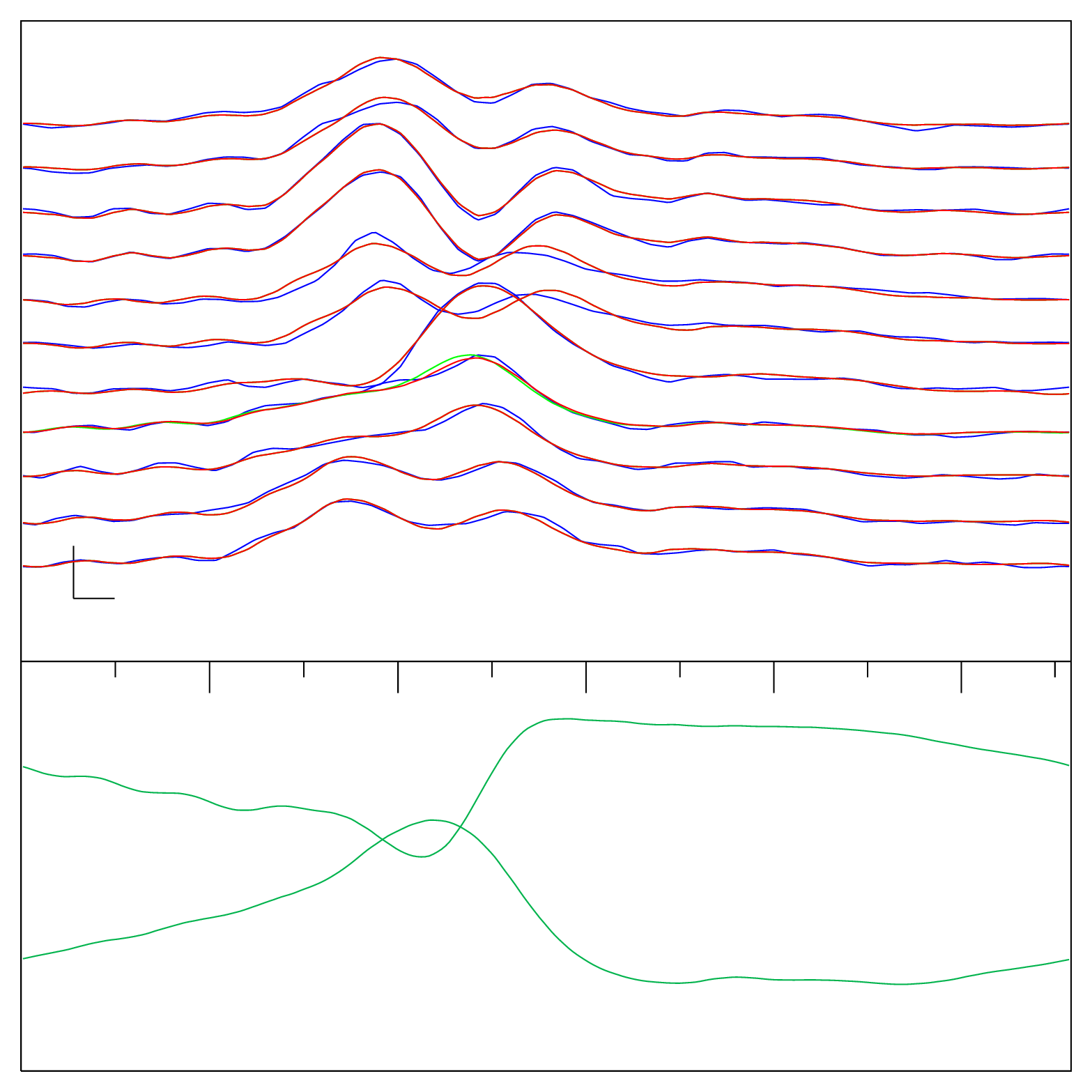}%{fc.eps}
\caption{Disentangled H$\alpha$ lines observed in high/soft state in
the 2004 (left panel) and low/hard state in 2006 (right panel). The
upper 7 lines (left panel) or 11 lines (right panel) indicate the
input spectra ordered in running time from the top. These profiles
are superimposed with their reconstruction as the sum of
disentangled components, which are shown by the two bottom curves.
The very bottom lines correspond to the emission of the focused
stellar wind, which is obviously higher in 2006 than in 2004. The
lines second from the bottom display the P-Cyg profiles of the
supergiant. The maxima of their emission wings are practically equal
in both states. } \label{figure:korelnew}
\end{figure}
\end{center}

\begin{center}
\begin{figure}[ht]
\centering
\includegraphics[bb=88 9 377 288, width=10cm]{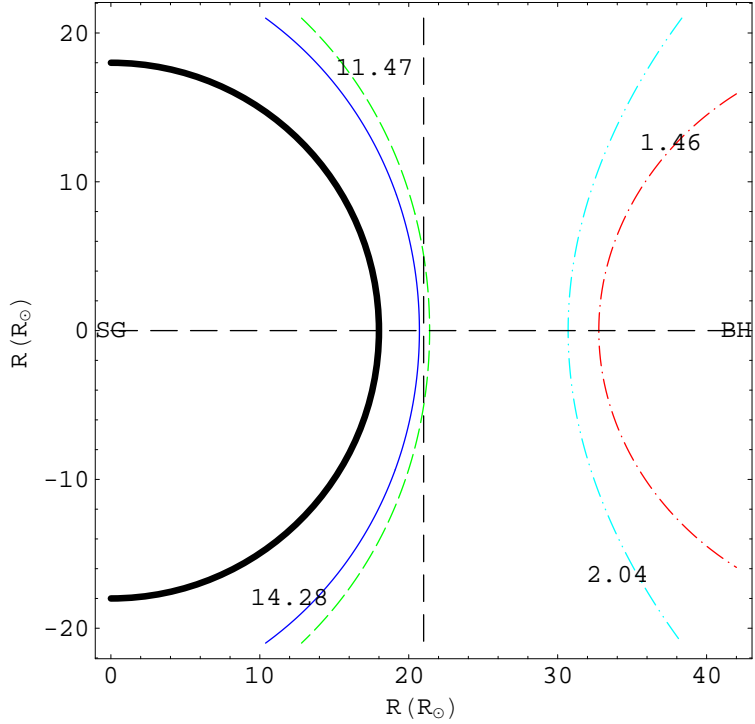}
\caption{Contours of $\xi$=10$^2$ for Cyg X-1 in different X-ray
states. The coordinate origin is at the center of the supergiant and
the thick circle represents the surface of the supergiant. The
intersection of the two black dashed lines is the boundary of the
supergiant's Roche lobe. The solid, dashed, dot-dashed, and
dot-dot-dashed lines are the contour lines when the X-ray luminosity
$(L_X)_{36}$ equals 14.28 (2001), 11.47 (2004), 1.46 (2003), and
2.04 (2006), respectively.} \label{figure:contour}
\end{figure}
\end{center}

\begin{center}
\begin{figure}[ht]
\centering
\includegraphics[bb=11 11 213 283, width=10cm]{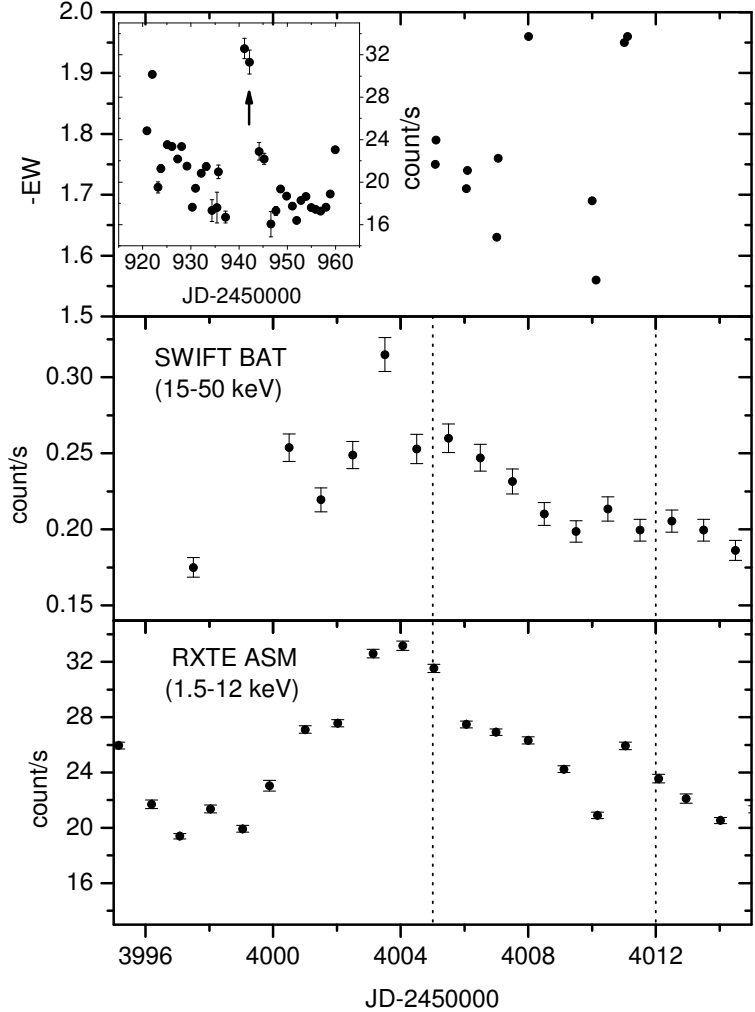}
\caption{The top panel shows the H$\alpha$ EWs during the 2006
observations. An X-ray flare was detected by Swift/BAT (15-50 keV)
(middle) and RXTE/ASM (1.5-12 keV) (bottom) around our observations.
The dashed lines correspond to the starting and ending times of the
2006 observations. The inset in the top panel is another X-ray flare
detected by RXTE/ASM (1.5-12 keV) around MJD 50,941.5449 (the
position of the arrow), when a strong H$\alpha$ emission
(--2.315\,{\AA}) was observed by \citet{tarasov2003}.}
\label{figure:flare}
\end{figure}
\end{center}

\acknowledgements J.Z.Y. is grateful to Min Fang for his help in
plotting Figure~\ref{figure:2006} and Figure~\ref{figure:contour}.
This research is partially supported by the National Natural Science
Foundation of China under grants 10433030 and 10673032. The work of
PH has been done in the framework of the Center for Theoretical
Astrophysics (ref.~LC06014) with a support of grant GA\v{C}R
202/06/0041. The authors appreciate valuable comments by the
referee.

\clearpage

\clearpage

\begin{deluxetable}{lccccccc}
\tablecolumns{9} \tablecaption{Summary of the spectroscopic observations of HDE 226868.}
\tablewidth{0pc}
\tablehead{\colhead{Date}&\colhead{UT} & \colhead{Exposure} &\colhead{Julian} & \colhead{Wavelength} & \colhead{Spectral} &\colhead{Orbital} & \colhead{H$\alpha$}\\
\colhead{}&\colhead{Middle} & \colhead{Time} &\colhead{Date} & \colhead{Range} & \colhead{Resolution} &\colhead{Phase} & \colhead{EW}\\
\colhead{}&\colhead{(hh:mm:ss)} & \colhead{(s)} &\colhead{} & \colhead{(\AA)} &
\colhead{\AA~pixel$^{-1}$} &\colhead{} & \colhead{(-\AA)}} \startdata
20010925   &  11:10:26  &  300  &  2452177.9656  &  5550-6750 & 1.22 & 0.916  &  0.33$\pm$0.02 \\
20010927   &  12:06:13  &  400  &  2452180.0043  &  4300-5500 & 1.22 & 0.280  &  --   \\
20010930   &  13:59:44  &  300  &  2452183.0832  &  5550-6750 & 1.22 & 0.830  &  0.11$\pm$0.01 \\
20011001   &  14:35:41  &  300  &  2452184.1081  &  5550-6750 & 1.22 & 0.013  &  0.80$\pm$0.02 \\
20021022   &  12:01:07  &  300  &  2452570.0008  &  5550-6750 & 1.22 & 0.925  &  1.11$\pm$0.14 \\
20021022   &  12:11:09  &  800  &  2452570.0077  &  5550-6750 & 1.22 & 0.926  &  1.24$\pm$0.09 \\
20021023   &  11:53:21  &  500  &  2452570.9954  &  5550-6750 & 1.22 & 0.102  &  1.10$\pm$0.11 \\
20021024   &  10:50:35  &  600  &  2452571.9518  &  5550-6750 & 1.22 & 0.273  &  1.19$\pm$0.08 \\
20021026   &  12:01:53  &  500  &  2452574.0013  &  5550-6750 & 1.22 & 0.639  &  1.04$\pm$0.05 \\
20021027   &  11:39:28  &  500  &  2452574.9857  &  4300-5500 & 1.22 & 0.815  &  --   \\
20021028   &  11:23:20  &  500  &  2452575.9745  &  5550-6750 & 1.22 & 0.991  &  0.85$\pm$0.11 \\
20031014   &  11:44:08  &  300  &  2452926.9890  &  5550-6750 & 1.22 & 0.674  &  1.36$\pm$0.04 \\
20031014   &  11:53:48  &  800  &  2452926.9957  &  5550-6750 & 1.22 & 0.676  &  1.54$\pm$0.04 \\
20031014   &  12:09:23  &  1000 &  2452927.0065  &  5550-6750 & 1.22 & 0.678  &  1.30$\pm$0.04 \\
20031015   &  11:48:38  &  300  &  2452927.9921  &  5550-6750 & 1.22 & 0.854  &  1.22$\pm$0.03 \\
20031015   &  11:56:49  &  500  &  2452927.9978  &  5550-6750 & 1.22 & 0.855  &  1.21$\pm$0.07 \\
20031015   &  12:05:49  &  500  &  2452928.0040  &  5550-6750 & 1.22 & 0.856  &  1.14$\pm$0.04 \\
20031016   &  11:44:54  &  20   &  2452928.9895  &  4300-6700 & 2.44 & 0.032  &  1.13$\pm$0.06 \\
20031016   &  11:46:34  &  100  &  2452928.9907  &  4300-6700 & 2.44 & 0.032  &  1.01$\pm$0.02 \\
20031016   &  11:48:20  &  50   &  2452928.9919  &  4300-6700 & 2.44 & 0.032  &  1.02$\pm$0.02 \\
20040921   &  14:53:34  &  900  &  2453270.1205  &  5550-6750 & 1.22 & 0.950  &  0.72$\pm$0.08 \\
20040921   &  15:09:20  &  900  &  2453270.1315  &  5550-6750 & 1.22 & 0.952  &  0.63$\pm$0.06 \\
20040922   &  14:43:41  &  900  &  2453271.1137  &  5550-6750 & 1.22 & 0.127  &  0.80$\pm$0.04 \\
20040923   &  11:35:11  &  600  &  2453271.9828  &  4300-6700 & 1.22 & 0.282  &  0.86$\pm$0.01 \\
20040925   &  13:55:31  &  1000 &  2453274.0802  &  5550-6750 & 1.22 & 0.657  &  0.44$\pm$0.01 \\
20040925   &  14:13:04  &  1000 &  2453274.0924  &  5550-6750 & 1.22 & 0.659  &  0.40$\pm$0.03 \\
20040926   &  13:31:28  &  1000 &  2453275.0635  &  5550-6750 & 1.22 & 0.832  &  0.28$\pm$0.01 \\
20040926   &  13:49:02  &  1000 &  2453275.0757  &  5550-6750 & 1.22 & 0.835  &  0.25$\pm$0.01 \\
20051021   &  13:30:33  &  900  &  2453665.0629  &  5550-6750 & 1.22 & 0.477  &  1.27$\pm$0.04 \\
20051023   &  11:31:50  &  100  &  2453666.9804  &  4300-6700 & 2.44 & 0.820  &  1.53$\pm$0.06 \\
20051023   &  11:34:05  &  100  &  2453666.9820  &  4300-6700 & 2.44 & 0.820  &  1.54$\pm$0.01 \\
20051024   &  12:17:10  &  900  &  2453668.0119  &  5550-6750 & 1.22 & 0.004  &  1.04$\pm$0.02 \\
20051024   &  12:32:50  &  900  &  2453668.0228  &  5550-6750 & 1.22 & 0.006  &  1.08$\pm$0.02 \\
20051027   &  12:56:50  &  600  &  2453671.0395  &  4300-5500 & 1.22 & 0.545  &  --   \\
20060926   &  14:05:43  &  1200 &  2454005.0873  &  5550-6750 & 1.02 & 0.198  &  1.75$\pm$0.07 \\
20060926   &  14:27:18  &  1200 &  2454005.1023  &  5550-6750 & 1.02 & 0.200  &  1.79$\pm$0.06 \\
20060927   &  13:26:20  &  1200 &  2454006.0600  &  5550-6750 & 1.02 & 0.372  &  1.71$\pm$0.02 \\
20060927   &  14:05:07  &  1200 &  2454006.0869  &  5550-6750 & 1.02 & 0.376  &  1.74$\pm$0.07 \\
20060928   &  12:24:02  &  1200 &  2454007.0167  &  5550-6750 & 1.02 & 0.542  &  1.63$\pm$0.09 \\
20060928   &  13:25:46  &  1200 &  2454007.0596  &  5550-6750 & 1.02 & 0.550  &  1.76$\pm$0.08 \\
20060929   &  12:19:54  &  1200 &  2454008.0138  &  5550-6750 & 1.02 & 0.720  &  1.96$\pm$0.03 \\
20060930   &  13:28:44  &  1200 &  2454009.0616  &  3850-5050 & 1.02 & 0.908  &  --   \\
20061001   &  12:04:59  &  600  &  2454010.0035  &  5550-6750 & 1.02 & 0.076  &  1.69$\pm$0.02 \\
20061001   &  15:09:09  &  600  &  2454010.1314  &  5550-6750 & 1.02 & 0.099  &  1.56$\pm$0.03 \\
20061002   &  12:09:59  &  600  &  2454011.0069  &  5550-6750 & 1.02 & 0.255  &  1.95$\pm$0.04 \\
20061002   &  14:35:00  &  600  &  2454011.1076  &  5550-6750 & 1.02 & 0.273  &  1.96$\pm$0.03 \\
\enddata
\tablenotetext{Note} {The ephemeris is adopted from Gies et al. (2003). } \label{table}
\end{deluxetable}

\begin{table*}[t]
\caption{Parameters of Cyg X-1}
\begin{tabular}{ccc} \hline \hline
  Parameter &  Value  & Reference\\ \hline
  $M_*$ & 24$\pm$5 $M_{\odot}$ & \citet{iorio2007}\\
  $M_x$ & 8.7$\pm$0.8 $M_{\odot}$ & \citet{shaposhnikov2007}\\
  q=$M_x$/$M_*$ & 0.36$\pm$0.05 & \citet{gies2003}\\
  $a$ & 42$\pm$9 $R_{\odot}$ & \citet{iorio2007}\\
  $i$ & 48.0$\pm$6.8$^{\circ}$ & \citet{iorio2007}\\
  Roche lobe size & $r_m$=21$\pm$6 $R_{\odot}$ &\citet{iorio2007}\\
  $V_{\infty}$ & 1700 km s$^{-1}$ & -\\
  $T_*$ & 40,000K & -\\
  $d$ & 2.5 kpc & \citet{ninkov87}  \\
  $L_{opt}/L_{\odot}$ & 5.59$\times$10$^{5}$ & \citet{ziolkowski2005}\\
  $\dot{M}_w$ & 2.0$\times$10$^{-6}$$M_{\odot}$ yr$^{-1}$ & \citet{howarth89}\\
  $R_*$ & 18 $R_{\odot}$ & -\\  \hline \hline
\end{tabular}
  \label{parameter}
\end{table*}

\begin{table*}[t]
  \caption{Ionization parameters in different X-ray states}
\begin{tabular}{ccccc} \hline \hline
  Observational & X-ray & X-ray luminosity & $\xi$($r_x$=21$R_{\odot}$) & $\xi$($r_x$=23$R_{\odot}$) \\
  Run           & State & (10$^{36}$ ergs s$^{-1}$) & (ergs cm s$^{-1}$)& (ergs cm s$^{-1}$)\\  \hline
  2001 & High/Soft & 14.28 & 110.04 & 40.89 \\
  2003 & Low/Hard  &  1.46 &  11.25 &  4.18 \\
  2004 & High/Soft & 11.47 &  88.38 & 32.84 \\
  2006 & Low/Hard  &  2.04 &  15.72 &  5.84 \\  \hline \hline
\end{tabular}
\label{ionization}
\end{table*}


\begin{thebibliography}{}
%\bibitem[Abbott(1982)]{abbott82}
%Abbott, D. C. 1982, ApJ, 259, 282
\bibitem[Albert et al.(2007)]{albert2007}
Albert, J., Aliu, E., Anderhub, H. et al. 2007, ApJ, 665 L51
\bibitem[Bagnuolo \& Gies(1991)]{bagnuolo91}
Bagnuolo, W. R., Jr., \& Gies, D. R. 1991, ApJ, 376, 266

\bibitem[Belloni et al.(1996)]{belloni96}
Belloni, T., M\'{e}ndez, M., van der Klis, M. et al. 1996, ApJ, 472, L107

\bibitem[Blondin et al.(1990)]{blondin90}
Blondin, J. M., Kallman, T. R., Fryxell, B. A., Taam, R. E. 1990, ApJ, 356, 591

\bibitem[Blondin et al.(1991)]{blondin91}
Blondin, J. M., Stevens, I. R., Kallman, T. R. 1991, ApJ, 371, 684

\bibitem[Bolton(1972)]{bolton72}
Bolton, C. T. 1972, Nature, 235, 271

\bibitem[Bowyer et al.(1965)]{bowyer65}
Bowyer, S., Byram, E. T., Chubb, T. A., Friedman, H. 1965, Sci, 147, 394

\bibitem[Brocksopp et al.(1999a)]{brocksopp99a}
Brocksopp, C., Fender, R. P., Larionov, V. et al. 1999a, MNRAS, 309,
1063

\bibitem[Brocksopp et al.(1999b)]{brocksopp99b}
Brocksopp, C., Tarasov, A. E., Lyuty, V. M., \& Roche, P. 1999b, A\&A, 343, 861



\bibitem[Castor, Abbott \& Klein(1975)]{castor75}
Castor, J. I., Abbott, D. C., Klein, R. I. 1975, ApJ, 195, 157

\bibitem[Done(2002)]{done2002}
Done, C. 2002, Phil. Trans. R. Soc. A, 360, 1967

\bibitem[Friend \& Abbott(1986)]{friend86}
Friend, D. B. \& Abbott, D. C. 1986, ApJ, 311, 701

\bibitem[Friend \& Castor(1982)]{friend82}
Friend, D. B. \& Castor, J. I. 1982, ApJ, 261, 293

\bibitem[Gallo et al.(2005)]{gallo2005}
Gallo, E., Fender, R., Kaiser, C. et al. 2005, Nature, 436, 819

\bibitem[Gies \& Bolton(1982)]{gies82}
Gies, D. R. \& Bolton, C. T. 1982, ApJ, 260, 240

\bibitem[Gies \& Bolton(1986a)]{gies86a}
Gies, D. R. \& Bolton, C. T. 1986a, ApJ, 304, 371

\bibitem[Gies \& Bolton(1986b)]{gies86b}
Gies, D. R. \& Bolton, C. T. 1986b, ApJ, 304, 389

\bibitem[Gies et al.(2003)]{gies2003}
Gies, D. R., Bolton, C. T., Thomson, J. R. et al. 2003, ApJ, 583, 424

\bibitem[Gies et al.(2008)]{gies2008}
Gies, D. R., Bolton, C. T., Blake, R. M. et al. 2008, ApJ, 678, 1237

\bibitem[Haberl et al.(1989)]{haberl89}
Haberl, F., White, N. E., Kallman, T. R. 1989, ApJ, 343, 409

\bibitem[Hadrava(1995)]{hadrava95}
Hadrava, P. 1995, A\&AS, 114, 393

\bibitem[Hadrava(1997)]{hadrava97}
Hadrava, P. 1997, A\&AS, 122, 581

\bibitem[Hadrava(2004)]{hadrava04}
Hadrava, P. 2004, Publ. Astron. Inst. ASCR, 92, 15

\bibitem[Hadrava(2006)]{hadrava06}
Hadrava, P. 2006, A\&AS, 448, 1149

\bibitem[Hadrava(2007)]{hadrava07}
Hadrava, P. 2007, in Proceedings of RAGtime 8/9: Workshops on black holes and neutron stars, ed. S.
Hled\'{\i}k \& Z. Stuchl\'{\i}k (Opava, Czech Republic: Silesian Univ.), 73 (arXiv:0710.0758)

\bibitem[Herrero et al.(1995)]{herrero95}
Herrero, A., Kudritzki, R. P., Gabler, R. et al. 1995, A\&A, 297, 556

\bibitem[Howarth \& Prinja(1989)]{howarth89}
Howarth, I. D. \& Prinja, R. K. 1989, ApJS, 69. 527

\bibitem[Iorio(2007)]{iorio2007}
Iorio, L. 2007, arXiv:0707.3525

\bibitem[Kallman \& McCray(1982)]{kallman82}
Kallman, T. R. \& McCray, R. 1982, ApJS, 50, 263

\bibitem[Kemp et al.(1983)]{kemp83}
Kemp, J. C., Barbour, M. S., Henson, G. D. et al. 1983, ApJ, 271, 65L

\bibitem[Lachowicz et al.(2006)]{lachowicz2006}
Lachowicz, P., Zdziarski, A. A., Schwarzenberg-Czerny, A. et al. 2006, MNRAS, 368, 1025L

\bibitem[LaSala et al.(1998)]{lasala98}
LaSala, J., Charles, P. A., Smith, R. A. D., Bauciska-Church, M., \& Church, M. J. 1998, MNRAS,
301, 285

\bibitem[Lasota(2001)]{lasota2001}
Lasota, J. P. 2001, New Astro. Rev., 45, 449

\bibitem[Malzac et al.(2006)]{malzac2006}
Malzac, J., Petrucci, P. O., Jourdain, E. et al. 2006, A\&A, 448, 1125

\bibitem[McClintock \& Remillard(2006)]{mcclintock2006}
McClintock, J. E. \& Remillard, R. A. 2006, in Compact Stellar X-ray Sources,
ed. W. H. G. Lewin, M. van der Klis %, pp. 157-214.
(Cambridge: Cambridge Univ. Press), 157 (astro-ph/0306213)


\bibitem[Miller et al.(2005)]{miller2005}
Miller, J. M., Wojdowski, P., Schulz, N. S. et al. 2005, ApJ, 620, 398

\bibitem[Mirabel \& Rodr\'{i}guez(1999)]{mirabel99}
Mirabel, I. F. \& Rodr\'{i}guez, L. F. 1999, ARA\&A, 37, 409

\bibitem[Ninkov et al.(1987)]{ninkov87}
Ninkov, Z., Walker, G. A. H. \& Yang, S. 1987, ApJ, 321, 438

\bibitem[Pauldrach et al.(1986)]{pauldrach86}
Pauldrach, A., Puls, J., Kudritzki, R. P. 1986, A\&A, 164, 86

\bibitem[Pooley(2001)]{pooley2001}
Pooley, G. 2001, IAUC, 7729, 3

\bibitem[Priedhorsky et al.(1983)]{priedhorsky83}
Priedhorsky, W. C., Terrell, J. \& Holt, S. S. 1983, ApJ, 270, 233

\bibitem[Puls et al.(1996)]{puls96}
Puls, J., Kudritzki, R.-P., Herrero, A. et al. 1996, A\&A, 305, 171

\bibitem[Remillard \& McClintock(2006)]{remillard2006}
Remillard, R. A. \& McClintock, J. E. 2006, ARA\&A, 44, 49

\bibitem[Richards \& Ratliff(1998)]{richards98}
Richards, M. T. \& Ratliff, M. A. 1998, ApJ, 565, 447

\bibitem[Russell et al.(2007)]{russell2007}
Russell, D. M., Fender, R. P., Gallo, E. \& Kaiser, C.R. 2007, MNRAS, 376, 1341

\bibitem[Sako et al.(1999)]{sako99}
Sako, M., Liedahl, D. A., Kahn, S. M. \& Paerels, F. 1999, ApJ, 525, 921

\bibitem[Shaposhnikov \& Titarchuk(2007)]{shaposhnikov2007}
Shaposhnikov, N. \& Titarchuk, L. 2007 ApJ, 663, 445

\bibitem[Sowers et al.(1998)]{sowers98}
Sowers, J. W., Gies, D. R., Bagnuolo, W. G. et al. 1998, ApJ, 506, 424

\bibitem[Stirling et al.(2001)]{stirling2001}
Stirling, A. M., Spencer, R. E., de la Force, C.J. et al. 2001, MNRAS, 327, 1273

\bibitem[Tanaka \& Lewin(1995)]{tanaka95}
Tanaka Y. \& Lewin W.H.G. 1995, in X-ray Binaries, ed. W. H. G. Lewin,
J. van Paradijs, E. P. J. van den Heuvel %, pp. 126-74.
(Cambridge: Cambridge Univ. Press), 126

\bibitem[Tarasov et al.(2003)]{tarasov2003}
Tarasov, A. E., Brocksopp, C., Lyuty, V. M. 2003, A\&A, 402, 237



\bibitem[Walborn(1973)]{walborn73}
Walborn, N. R. 1973, ApJ, 179, L123

\bibitem[Webster \& Murdin(1972)]{webster72}
Webster, B. L. \& Murdin, P. 1972, Nature, 235, 37

\bibitem[Wijers \& Pringle(1999)]{wijers99}
Wijers, R. A. M. J. \& Pringle, J. E. 1999, MNRAS, 308, 207

\bibitem[Wu et al.(2001)]{wu2001}
Wu, K., Soria, R., Hunstead, R. W., Johnston, H. M. 2001, MNRAS, 320, 177

\bibitem[Yan, Liu \& Hang(2005)]{yan2005}
Yan, J. Z., Liu, Q. Z. \& Hang, H. R. 2005, ChJAA, 5S, 247

\bibitem[Zdziarski et al.(2002)]{zdziarski2002}
Zdziarski, A. A., Poutanen, J., Paciesas, W. S., Wen, L. Q. 2002, ApJ, 578, 357

\bibitem[Zhang et al.(1997)]{zhang97}
Zhang, S. N., Cui, W, Harmon, B. A. et al. 1997, ApJ, 477, L95

\bibitem[Zi\'{o}{\l}kowski(2005)]{ziolkowski2005}
Zi\'{o}{\l}kowski, J. 2005, MNRAS, 358, 851

\end{thebibliography}
\end{document}